\documentclass[manuscript]{aastex}

\shorttitle{N$_2$ and CO trapping in clathrates}
\shortauthors{Lectez et al.}

\usepackage{etex}
\usepackage{m-pictex, m-ch-en}
\usepackage {amsmath}

\begin{document}


\title{A $\sim$32-70 K formation temperature range for the ice grains agglomerated by comet 67P/Churyumov-Gerasimenko}


\author{S. Lectez\altaffilmark{1}, J.-M. Simon\altaffilmark{1}, O. Mousis\altaffilmark{2}, S. Picaud\altaffilmark{3}, K. Altwegg\altaffilmark{4}, M. Rubin\altaffilmark{4}, J.M. Salazar\altaffilmark{1}}


\altaffiltext{1}{Laboratoire Interdisciplinaire Carnot de Bourgogne, UMR 6303 CNRS-Universit\'e de Bourgogne, Dijon, France {\tt jmsimon@u-bourgogne.fr}}
\altaffiltext{2}{Aix Marseille Universit{\'e}, CNRS, LAM (Laboratoire d'Astrophysique de Marseille) UMR 7326, 13388, Marseille, France}
\altaffiltext{3}{Institut UTINAM, UMR 6213 CNRS-Universit\'e de Franche Comt\'e, Besan\c con, France}
\altaffiltext{4}{Physikalisches Institut, University of Bern, Sidlerstrasse 5, CH-3012 Bern, Switzerland}


\begin{abstract}
Grand Canonical Monte Carlo simulations are used to reproduce the N$_2$/CO ratio ranging between 1.7 ~$\times$~10$^{-3}$ and 1.6 ~$\times$~10$^{-2}$ observed {\it in situ} in the Jupiter family comet 67P/Churyumov-Gerasimenko by the ROSINA mass spectrometer aboard the Rosetta spacecraft, assuming that this body has been agglomerated from clathrates in the protosolar nebula. Simulations are done using an elaborated interatomic potentials for investigating the temperature dependence of the trapping within a multiple guest clathrate formed from a gas mixture of CO and N$_2$ in proportions corresponding to those expected for the protosolar nebula. By assuming that 67P/Churyumov-Gerasimenko agglomerated from clathrates, our calculations suggest the cometary grains must have been formed at temperatures ranging between $\sim$31.8 and 69.9~K in the protosolar nebula to match the N$_2$/CO ratio measured by the ROSINA mass spectrometer. The presence of clathrates in Jupiter family comets could then explain the potential N$_2$ depletion  (factor up to $\sim$87 compared to the protosolar value) measured in 67P/Churyumov-Gerasimenko.\\
\end{abstract}


\section{Introduction}

The thermodynamic conditions prevailing in many bodies of the solar system suggest that clathrates could exist in the Martian permafrost \citep{T09,S09,M13}, on Titan \citep{CS12,M14}, as well as in the interiors of other icy satellites \citep{K06,H06}. It has also been suggested that the activity observed in some cometary nuclei results from the dissociation of these crystalline structures \citep{Mar83,S88,Ma10,Ma11,Ma12,M12}. Several indirect evidences suggest that clathrates probably participated in the formation of planetesimals and the building blocks of giant planets in the outer solar system \citep{LS85,M10,M14}. 

In the absence of experimental data existing at low-temperature (20--200 K) and low-pressure conditions (10$^{-13}$--10$^{-3}$ bar), which are typical of those encountered in planetary environments \citep{LS85,S08}, clathrates are characterized via theoretical modeling. The approach usually employed in planetary science is based on the statistical mechanics model initially proposed by \cite{vdWP}, which makes use of simplified intermolecular potentials calibrated on equilibrium measurements performed at relatively high temperatures. In consequence, the use of these potentials at thermodynamic conditions relevant to those of the protosolar nebula (hereafter PSN) for predicting the composition of clathrates deserves to be confronted with more sophisticated approaches.

In this present work,  we aim at reproducing the N$_2$/CO ratio  ranging between 1.7 ~$\times$~10$^{-3}$ and 1.6 ~$\times$~10$^{-2}$ observed {\it in situ} in the Jupiter family comet 67P/Churyumov-Gerasimenko (hereafter 67P) by the ROSINA mass spectrometer aboard the Rosetta spacecraft \citep{B07}, which is found to be depleted by a factor up to $\sim$87 compared to the value of 0.148 hypothesized for the protosolar nebula (see below). By assuming that 67P has been agglomerated from clathrates, it is possible to derive the temperature range of formation of these crystalline structures in the PSN by mean of Grand Canonical Monte Carlo (GCMC) simulations based on elaborated interatomic potentials. These allowed us to investigate the temperature dependence of the trapping within a multiple guest (hereafter MG) clathrate formed from a gaseous mixture of CO and N$_2$ in proportions corresponding to those expected for the protosolar nebula.

\section{Computational details and simulation procedure}

We assume multiple guest (MG) clathrate formation from a gaseous mixture composed of N$_2$ and CO in proportions specified from a plausible protosolar gas phase composition. For this, we have assumed that both all C and N are in form of CO and N$_2$ in the initial disk's gas phase. This assumption is in agreement with thermochemical models of the protosolar nebula \citep{L80,P89,M02}. The use of C and N protosolar elemental abundances from the compilation of \cite{L09} allowed us to derive a gaseous mixture with mole fractions of 0.871 for CO and 0.129 for N$_2$, giving a N$_2$/CO ratio of 0.148 in the PSN.

The MG clathrate composition has been computed along its equilibrium pressure curve $P_{{\text{eq},MG}}$ given by \citep{T07,M08}:

\begin{equation}
 P_{{\text{eq},MG}} = \left [\sum_{{\text{i}}} \frac{y_i} {P_{{\text{eq},i}}} \right]^{-1},
\label{eq1}
\end{equation}

\noindent where $y_i$ is the mole fraction of species $i$ in the gas phase (here CO or N$_2$) and $P_{{\text{eq},i}}$ the equilibrium pressure of single guest clathrate formed from component $i$ only.

\begin{figure}[h]
\begin{center}
\resizebox{\hsize}{!}{\includegraphics[angle=0]{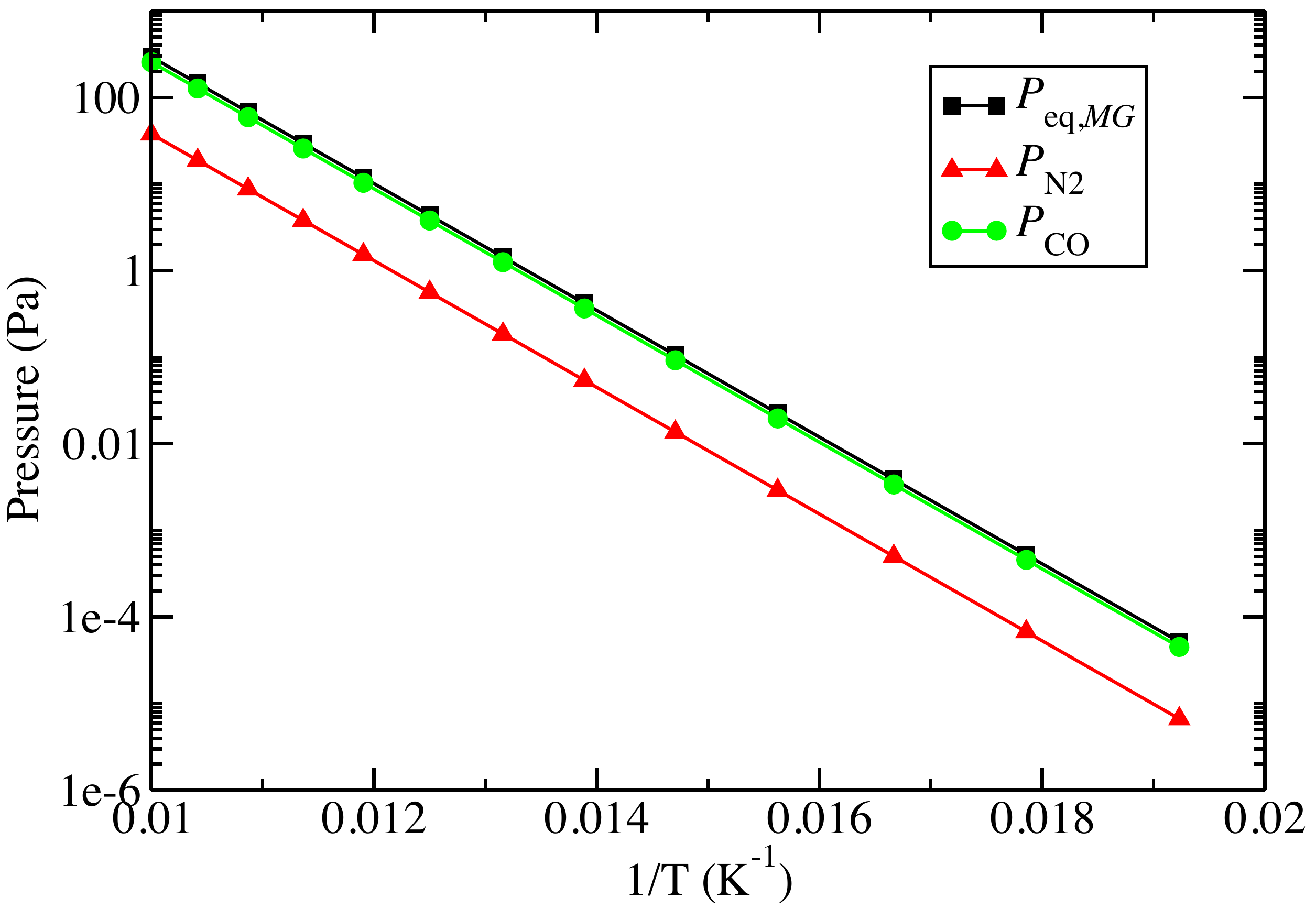}}
\caption{Calculated pressure and partial pressures of CO and N$_2$ in the gas as a function of the inverse temperature based on the Arrhenius law and using parameters in Table \ref{param}.}
\label{fig1}
\end{center}
\end{figure}

The equilibrium pressures $P_{{\text{eq},i}}$ (in bar) are determined by using an equation based on an Arrhenius law \citep{M61}:

\begin{equation}
ln~P_{\text{eq},i} = \frac{A_i} {T} + B_i,
\label{eq2}
\end{equation}

\noindent where $A_i$ and $B_i$ are constant parameters depending on the nature of the species trapped in the clathrate and $T$ is the temperature (K). $A_i$ and $B_i$ have been determined by fitting the available theoretical and laboratory data \citep{MM08} and their values are given in Table \ref{param}. The calculated values of $P_{{\text{eq},i}}$ and partial pressures of CO ($P_{\rm CO}$) and N$_2$ ($P_{\rm N_2}$) are represented as a function of the inverse temperature in Fig. \ref{fig1}. Here, the equilibrium pressure of MG clathrate is in the $\sim$5.2 $\times$10$^{-10}$--2.9 $\times$10$^{-3}$ bar range when $T$ is varied between $\sim$52 and 100~K, respectively. 

We have calculated the composition of N$_2$--CO clathrates via Monte-carlo (MC) simulations in the Grand Canonical ensemble (GCMC)  \citep{Fren2002} for temperatures ranging from  52 to 100~K (the choice of this temperature range is explained in Sec. \ref{sec:results}, but it is worth noting that, below 50 K, the equilibration time is in fact too long for the GCMC simulation), with an increment of 4~K between each computation. All our calculations have been performed in the case of Structure I (SI) clathrates (see structural details in \cite{S08}). This choose has been motivated by the fact that CO is the dominating species in the gaseous mixture and is known to form Structure I single guest clathrate \citep{M05}.

In our system, the considered crystal size consists in 125 cubic unit cells (5$\times$5$\times$5), corresponding to 5,750 water molecules. The dimension of one parameter of the cubic simulation box is set equal to 60.15 \AA~during all our simulations. Periodic boundary conditions are applied to mimic infinite crystal. The water molecules are modeled using the well-known TIP4P/2005 model \citep{Abascal2005}, allowing them to translate and rotate during the simulation. Models for N$_2$ and CO molecules are taken from \cite{Potoff2001} and \cite{Piper1984}, respectively. One hundred million MC steps were performed including insertion, deletion, translation and rotation of the molecules. Only the last 50 million steps were used to compute the data. The first 50 million steps have been discarded from the analysis and were only used to equilibrate the system.

From the knowledge of the partial pressure of each species (see Fig. \ref{fig1}), GCMC simulations have been performed to compute the composition of the MG clathrate. However, the number of MC steps needed to equilibrate the system strongly increases  when temperature decreases. Simulation tests showed that below a temperature of 52~K, this number becomes even larger than the 50 million steps we simulated. At these low temperatures, statistically relevant results on the mole fractions of encaged molecules in our MC calculations thus become too time consuming. In consequence, below 52~K, these quantities have been evaluated via a thermodynamic extrapolation, described below, and fitted to the results computed at higher temperatures.

After the equilibration period, chemical equilibrium takes place between gas and clathrate. It is then possible to write an equilibrium constant $K_i$ for each species $i$ in the form:

\begin{equation}
K_i = \frac {a_i}{P_i/P^0},
\label{eq3}
\end{equation}

\noindent where $a_i$ is the activity of species $i$ in clathrate defined from the mole fraction of species $i$ in clathrate ($x_i$) and its activity coefficient ($\gamma_i$ ) as
$a_i = \gamma_i x_i$. The evolution of $K_i$ with temperature should obey the Van't Hoff relation:

\begin{equation}
\frac{\text{dln}K_i}{\text{d}T} = \frac {\Delta E_i}{{\text R}T^2} 
\label{eq4},
\end{equation}

\noindent where $\Delta E_i$ is the entrapping energy of CO or N$_2$, and R the ideal gas constant. 

\begin{figure}[h]
\begin{center}
\resizebox{\hsize}{!}{\includegraphics[angle=0]{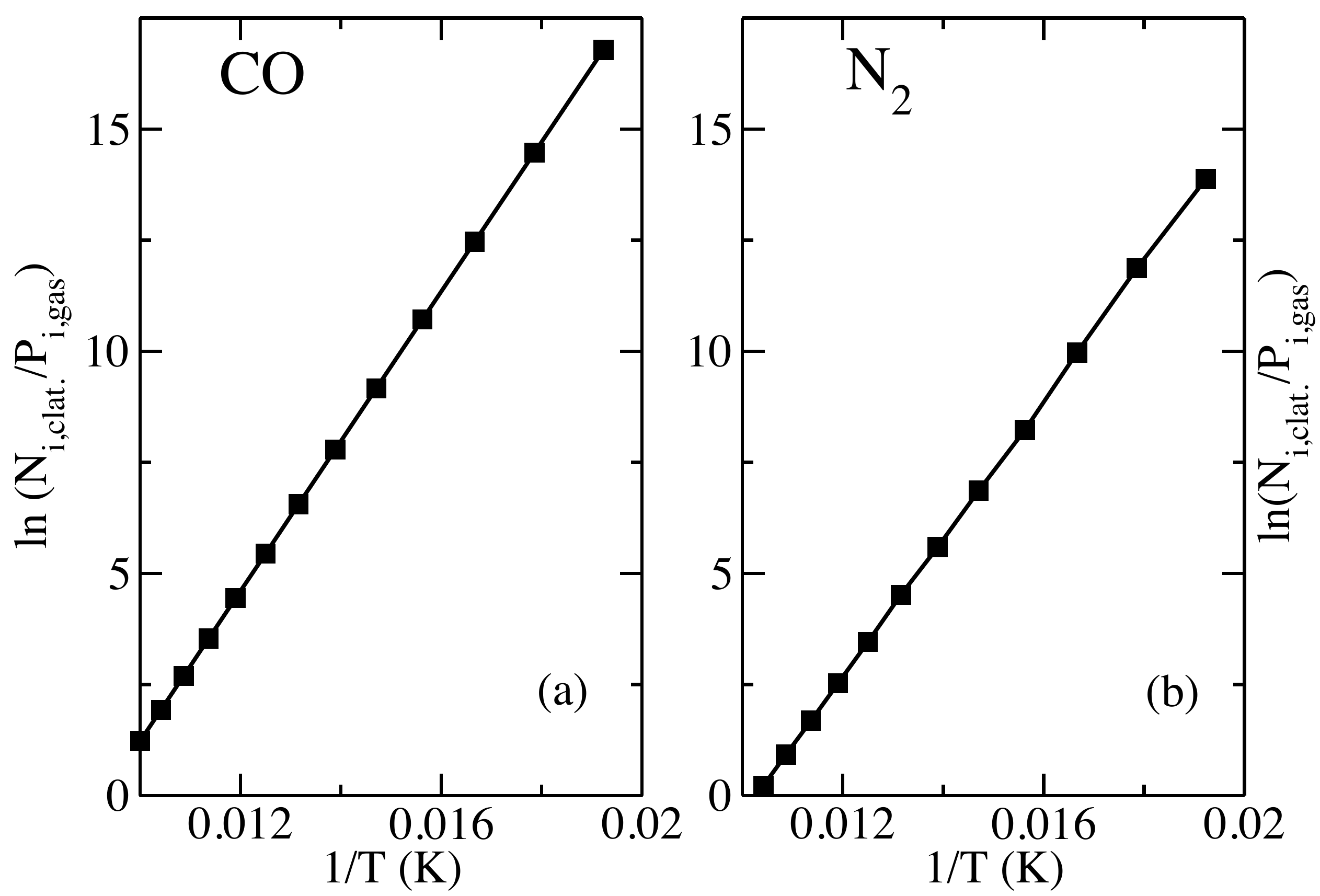}}
\caption{Evolution of the number of encaged molecules in the cases of CO (a) and N$_2$ (b) as a function of the inverse temperature. A linear fit was performed on the data with correlation coefficients higher than 0.999 in both cases (see text).  The figure shows the results of the GCMC simulation between 52K and 100K.}
\label{fig2}
\end{center}
\end{figure}

Under the simulated conditions, the occupancy of the clathrate cages is still maximum. We thus, obtained an amount of 8 molecules of CO and N$_2$ per unit cell, implying that the total number of molecules trapped in clathrate is constant, independently of the temperature. By Neglecting the non ideal terms of the activity coefficient at first approximation, this allowed us to get $\gamma_i~=~1$ and to rewrite Eq. \ref{eq4} as follows:

\begin{equation}
\frac{\text{dln}(N_i/P_i)}{\text{d}(1/T)} = -\frac{\Delta E_i}{\text R} 
\label{eq5},
\end{equation}

\noindent with $N_i$ the number of encaged molecules of type $i$. Our simulations allowed retrieving the values of $N_{\rm CO}$ and $N_{\rm N_2}$ at temperatures higher than 52~K and provided us with a direct access to the different values of the quantity ln$(N_i/P_i)$. Figure~\ref{fig2} represents these quantities plotted for CO and N$_2$ as a function of $1/T$. They show a linear behavior with a correlation coefficient higher than 0.999 for each species, in agreement with Eq. \ref{eq5}. The computed data have been fitted via the use of linear equations, allowing to find ln($N_{\rm CO}/P_{\rm CO}$)~=~1685.145/$T$~-~15.617 and ln($N_{\rm N_2}/P_{\rm N_2}$)~=~1554.469/$T$~-~15.972. The entrapping energies comes directly from the fit, $\Delta E_{\rm CO}=-1685.145\times$R and $\Delta E_{\rm N2}=-1554.469\times$R. These two linear equations have been used to estimate the $N_{\rm N2}/N_{\rm CO}$  ratio at temperatures lower than 52~K. Note that, for simplification, we will use below the abbreviation N$_2$/CO for this ratio. 

\section{Results}
\label{sec:results}

\begin{figure}[h]
\begin{center}
\resizebox{\hsize}{!}{\includegraphics[angle=0]{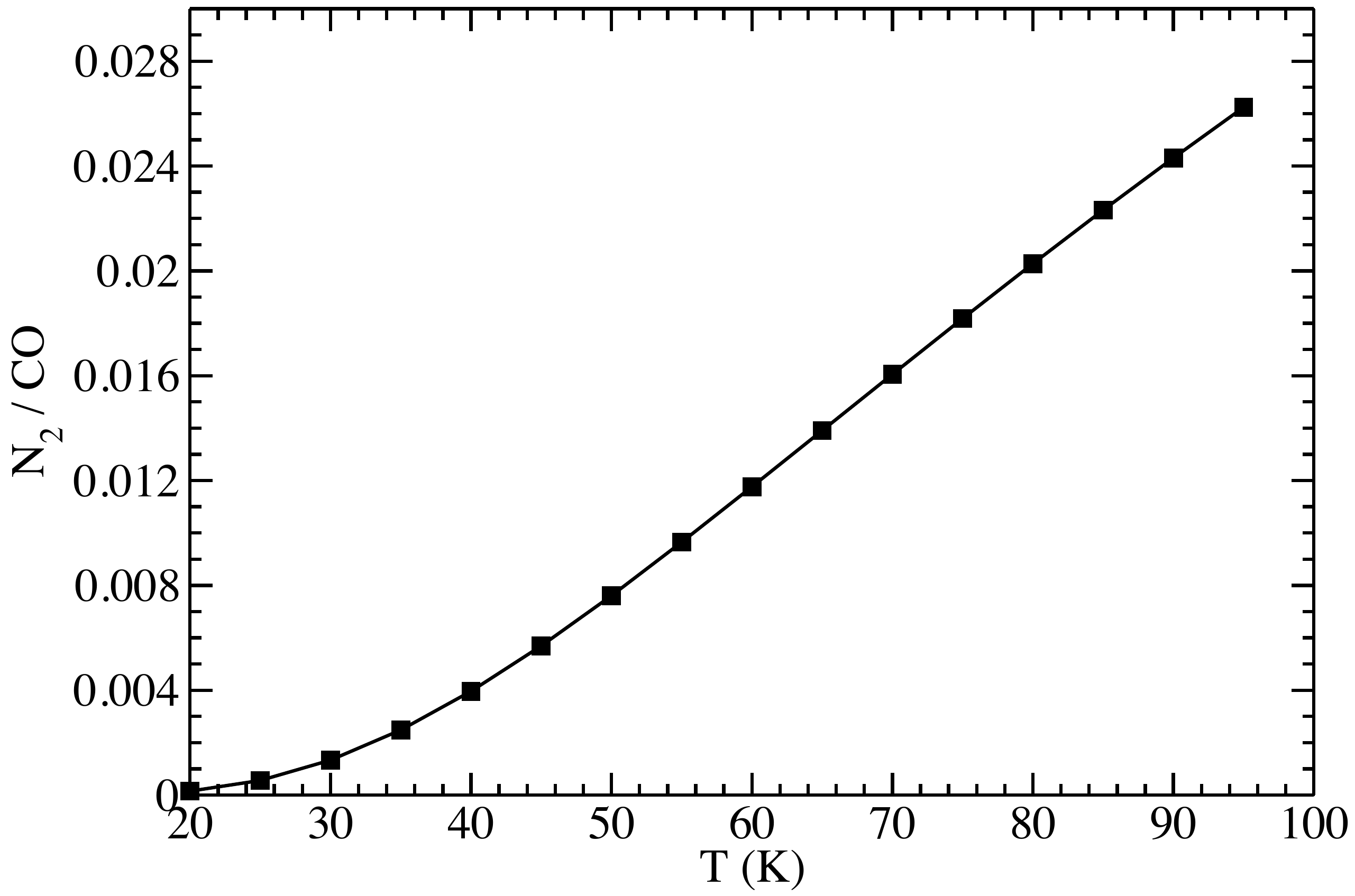}}
\caption{N$_2$/CO ratio in clathrate as a function of formation temperature. The results are derived from the GCMC simulations performed above 52K. Below this temperature, they are based on the entrapping energies derived from the GCMC simulations.}
\label{fig3}
\end{center}
\end{figure}

Figure~\ref{fig3} shows the evolution of  N$_2$/CO ratio as a function of temperature in the 20--100~K range. This ratio monotonically increases with the growing temperature. The figure exhibits a linear regime in the 50--100~K range whereas at lower temperatures, the curve trend is less steep and the ratio converges smoothly towards zero. The N$_2$/CO ratio found is equal to $\sim$1.5~$\times$~10$^{-4}$ at 20~K. This value is 50 times smaller than at 50~K. In the temperature range considered here, the calculated N$_2$/CO ratio is significantly lower than the ratio of $\sim$0.15 in the coexisting gas phase. This indicates that the clathrate formation favors the CO entrapping at  the expense of N$_2$. This behavior is related to  the difference in entrapping energy between CO and  N$_2$. Indeed, $\Delta E_{\rm CO}$  being lower, the entrapping of CO is selectively favored when the temperature decreases.

\begin{figure}[h]
\begin{center}
\resizebox{\hsize}{!}{\includegraphics[angle=0]{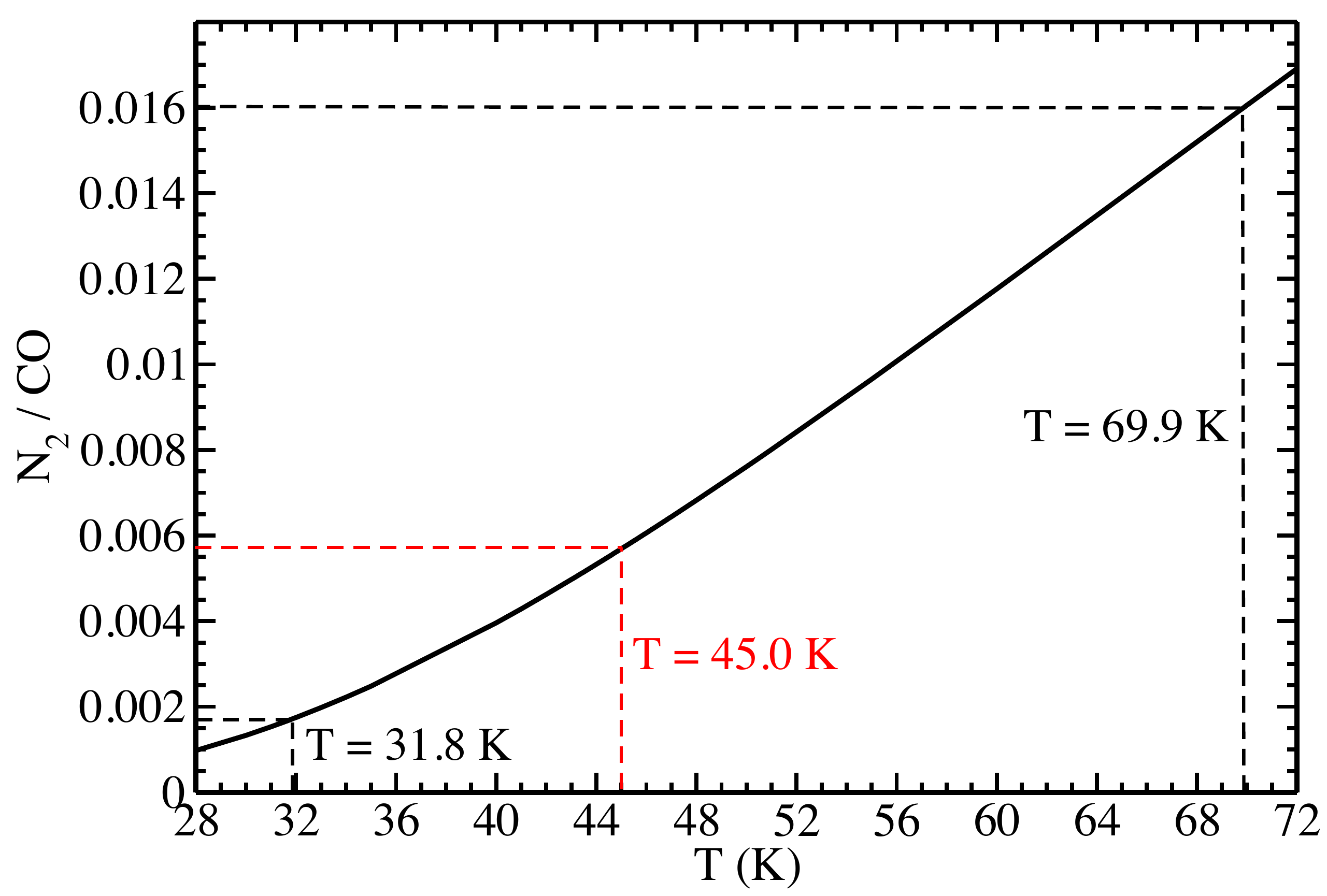}}
\caption{ Minimum and maximum N$_2$/CO ratios measured in 67P and corresponding formation temperatures for the ice grains. The results are derived from the GCMC simulations performed above 52K. Below this temperature, they are based on the entrapping energies derived from the GCMC simulations.}
\label{fig4}
\end{center}
\end{figure}

Figure~\ref{fig4} is a zoom of a portion of Fig.~\ref{fig3} given for easy reading of the correspondence between the N$_2$/CO ratio measured in 67P by the ROSINA instrument and the formation temperature of the ice grains from which the comet agglomerated. Taking into account the strong variation of the N$_2$/CO measurement between 0.17 to 1.6\% depending on the position of the Rosetta spacecraft above the surface of the comet nucleus \citep{R15}, we find that the ice grains at the origin of 67P formed at temperatures ranging between $\sim$31.8 and 69.9 K in the protosolar nebula, with corresponding equilibrium pressures ranging between 6.0 $\times$ 10$^{-19}$ and 2.1 $\times$ 10$^{-6}$ bar. For the sake of information, the mean N$_2$/CO ratio of 0.57\% corresponding to the averaging of the 138 spectra obtained by \cite{R15} is represented on Fig. \ref{fig4}. The corresponding formation temperature of the ice grains is of $\sim$45 K in the PSN, with an equilibrium pressure of 3.3~$\times$~10$^{-12}$ bar.

\section{Discussion and Conclusions}

The composition of a MG clathrate formed from a gaseous mixture of N$_2$ and CO in proportions corresponding to those expected for the protosolar nebula (87.1 \% for CO and 12.9\% for N$_2$) has been investigated in the 20--100 K temperature range. Above 50 K, the clathrate composition has been computed via Grand Canonical Monte-Carlo simulations for pressures ranging from $\sim$5.2 $\times$10$^{-10}$ to 2.9 $\times$10$^{-3}$ bar. Below 50 K, the clathrate composition has been extrapolated via the use of a Van't Hoff relation. The results show that, at thermodynamic conditions relevant to those of the protosolar nebula, CO has a much higher propensity than N$_2$ to be trapped in clathrates. Assuming that 67P agglomerated from clathrates, our calculations suggest that the cometary grains must have formed  at temperatures ranging between $\sim$31.8 and 69.9 K in the protosolar nebula to match the N$_2$/CO ratio measured by the ROSINA mass spectrometer \citep{R15}. 

Whilst narrower, the range of formation temperatures inferred from our model for the grains of 67P is consistent with the one ($\sim$22--80 K) found from the reading of Fig. 2 of \cite{M12} who performed calculations of planetesimals compositions based on the classical statistical mechanics model of \cite{vdWP}. In the absence of experiments at this temperature range, the fact that these two different approaches lead to similar conclusions, namely that clathrates can explain the N$_2$/CO ratio observed in 67P, suggest that this scenario is plausible. The presence of clathrates in Jupiter family comets could then explain the apparent N$_2$ depletion (factor up to $\sim$87 compared to the protosolar value) measured in 67P.

\acknowledgements
Financial support from the BQR Bourgogne Franche--Comt{\'e} is gratefully acknowledged. O.M. acknowledges support from CNES. This work has been partly carried out thanks to the support of the A*MIDEX project (n\textsuperscript{o} ANR-11-IDEX-0001-02) funded by the ``Investissements d'Avenir'' French Government program, managed by the French National Research Agency (ANR).

\clearpage

\begin{table}[h]
\centering 
\caption{Parameters of the equilibrium curves of the considered single guest clathrates}
\begin{tabular}{lcc}
\hline 
\hline
Molecule type	$i$				& $A_i$ / (K)		&  $B_i$		\\
\hline
CO             					& -1685.54	& 10.9946    	\\
N$_2$              					& -1677.62	& 11.1919    	\\
\hline
\end{tabular}
\label{param}
\end{table}

\clearpage

\clearpage

\clearpage

\clearpage

\end{document}